\documentclass[twoside]{article}
\pdfoutput=1
\usepackage[pdftex]{graphicx}
\usepackage{aas_macros}
\usepackage{amsmath}
\usepackage{natbib}
\usepackage{deluxetable}
\usepackage{setspace}
\usepackage{multirow}
\usepackage{rotating}
\usepackage{hyperref}

\newenvironment{thinfigure}
  {\par\medskip\noindent\minipage{\linewidth}}
  {\endminipage\par\medskip}

\usepackage[sc]{mathpazo} 
\usepackage[T1]{fontenc} 
\usepackage{microtype} 

\usepackage[hmarginratio=1:1,top=32mm,columnsep=20pt]{geometry} 
\usepackage{multicol} 
\usepackage[hang, small,labelfont=bf,up,textfont=it,up]{caption} 
\usepackage{booktabs} 
\usepackage{float} 

\usepackage{lettrine} 
\usepackage{paralist} 

\usepackage{abstract} 

\usepackage{titlesec} 
\titleformat{\section}[block]{\large\scshape\centering}{\thesection.}{1em}{} 
\titleformat{\subsection}[block]{\small}{\thesubsection.}{1em}{} 

\usepackage{fancyhdr} 
\title{\vspace{-15mm}\fontsize{24pt}{10pt}\selectfont\textbf{Quantitative Comparison of Methods for Predicting the Arrival of Coronal Mass Ejections at Earth based on multi-view imaging}} 

\author{
\large
\textsc{R. C. Colaninno, A. Vourlidas, C.-C. Wu }\\[2mm] 
\normalsize Space Science Division, Naval Research Laboratory, Washington, District of Columbia, USA.\\ 
\normalsize \href{mailto:robin.colaninno@nrl.navy.mil}{robin.colaninno@nrl.navy.mil} 
\vspace{-5mm}
}
\date{}

\begin{document}

\maketitle 

\thispagestyle{fancy} 


\begin{abstract}

We investigate the performance of six methods for predicting the CME time of arrival (ToA) and velocity at Earth using a sample of nine Earth-impacting CMEs between May 2010 and June 2011. The CMEs were tracked continuously from the Sun to near Earth in multi-viewpoint imaging data from STEREO SECCHI and SOHO LASCO. We use the Graduate Cylindrical Shell (GCS) model to estimate the three-dimensional direction and height of the CMEs in every image out to $\sim$200 R$_\odot$. We fit the derived three-dimensional (deprojected) height and time data with six different methods to extrapolate the CME ToA and velocity at Earth. We compare the fitting results with the in situ data from the WIND spacecraft. We find that a simple linear fit after a height of 50$R_\odot$ gives the best ToA with a total error $\pm$13 hours. For seven (78\%) of the CMEs, we are able to predict the ToA to within $\pm$6 hours. These results are a full day improvement over past CME arrival time methods that only used SOHO LASCO data. We conclude that heliographic measurements, beyond the coronagraphic field of view, of the CME front made away from the Sun-Earth line are essential for accurate predictions of their time of arrival.    

\end{abstract}


\begin{multicols}{2} 

\section{Introduction}
In the last few decades, we have discovered that the space environment around our planet is as dynamic as terrestrial weather. The source of this space weather is the Sun which produces winds and storms that affect modern infrastructure. The most geo-effective aspects of space weather are coronal mass ejections (CMEs) which are analogous to terrestrial hurricanes.  These powerful storms, comprised of plasma and magnetic fields ejected from the solar corona, can significantly disrupt Earth's magnetosphere and cause a range of terrestrial effects from the aurora to ground induced currents. CMEs were first observed in visible-light coronagraphs  \citep{1972BAAS....4R.394T} as bright large-scale density enhancements propagating outwards from the Sun. Signatures of CMEs are also seen \emph{in situ} plasma and magnetic field data \citep{2000GeoRL..27.3591C}. \emph{In situ} measurements at Earth give us the real-time arrival and physical properties of geo-effective CMEs. Throughout the paper, we use the term CME to describe both the imaging and \emph{in situ} observations.

Since mid-2007, we are able to continuously monitor the propagation of CMEs from the Sun to Earth using the observations from the Sun-Earth Connection Coronal and Heliospheric Investigation (SECCHI; \citealt{2008SSRv..136...67H}) instrument suite aboard the \emph{Solar TErrestrial RElations Observatory} mission  (\emph{STEREO}; \citealt{2008SSRv..136....5K}).  However, even with complete coverage of the Sun-Earth line with visible-light imaging data, it remains difficult to accurately predict the arrival of CMEs at Earth.  CMEs are detectable in visible-light due to Thomson scattering of photospheric sunlight by the electrons within the CME. This emission is optically thin, thus, the observations are integrations along the line of sight (LOS) making it difficult to identify individual features of the diffuse CME structure. The emission from the CME electrons drops off quickly as the CME expands away from the Sun causing a decrease in density and scattering efficiency with the viewing angle. By the time CMEs reach Earth, they are large ($\sim$ 0.5~AU width) with correspondingly long LOS integration paths. All these elements of CME observations at large heliocentric distances make it difficult to know the time of arrival (ToA) even when both the Earth and the CME are visible simultaneously in the same field of view (FOV).

With careful treatment of the visible-light image data, the position of the CME front can be measured and the resulting height and time (HT) data points can be fitted by a curve (e.g. polynomial or spline) to model its motion. If the height of the CME is not measured up to 1~AU, the fit can then be extrapolated to predict the ToA of the CME at Earth. This is one of the most basic space weather prediction techniques that can be applied to imaging observations. 

Even prior to the availability of heliographic data from SECCHI, several methods had been proposed in the literature to predict the ToA using coronagraphic data from the Large Angle and Spectroscopic Coronagraph (LASCO; \citealt{1995SoPh..162..357B}) aboard the \emph{SOlar and Heliospheric Observatory} mission (\emph{SOHO}; \citealt{1995SoPh..162....1D}). Empirical CME propagation models \citep{2000GeoRL..27..145G,2001JGR...10629207G, 2005AnGeo..23.1033S} were developed using the LASCO data with limited success. These models use the projected velocity of Earth-directed CMEs observed in the LASCO FOV.  \citet{2004AnGeo..22..661O} evaluated the predicted ToA for three empirical models: \citet{2000GeoRL..27..145G,2001JGR...10629207G} and \citet{2002JGRA..107.1019V}. They found little difference between the three models with an average ToA error of 0.46 days. \citet{2005AnGeo..23.1033S} reported similar results for their method which used the expansion speed of the CME as a proxy for the CME radial speed. All these models can predict the arrival of the CME at Earth within a $\pm$24 hours window with a 95\%  {error margin}. 

At the time, the inaccuracy of these models was attributed to the inability to measure the true (deprojected) speed of Earth-directed CMEs since LASCO is located along the Sun-Earth line offering a head-on view of the propagation. \citet{1999JGR...10412515L} compiled a data set of CMEs observed in quadrature using Solar Maximum Mission and Solwind coronagraphs, and \emph{Helios~1} and \emph{Pioneer Venus Orbiter} \emph{in situ} magnetic field and plasma measurements. \citet{2004AnGeo..22..661O} found that using these data where the projection effects are minimized did not improve the results of the three studied ToA models. Physics-based shock propagation models, using the speeds from the metric type-II radio bursts which are not affected by projection effects, do not fare any better with ToA predictions \citep{2003JGRA..108.1445C}. 

\citet{2012SoPh..279..477K} fitted the stereoscopic data from SECCHI and LASCO with the Graduated Cylindrical Shell (GCS;  \citealt{2009SoPh..256..111T}) model to derive the three-dimensional (3D) position, direction and speed of 30 CMEs within the LASCO FOV ($<$~30 R$_\odot$) between 2008 and 2010.  {\citet{2012SoPh..279..477K} did not extend their analysis into the SECCHI HI FOV.} They applied these deprojected CME speeds in the models of \citet{2000GeoRL..27..145G, 2001JGR...10629207G}. They compared the predicted ToA to the \emph{in situ} measurements and found an error of $\pm$30 hours with a 95\%  {error margin}. This result is actually worse than the predictions obtained using the same model with projected CME speeds. It is unclear what caused this unexpected result. It may be due to the low average speeds of the \citet{2012SoPh..279..477K} sample,  { taken during the most unusual minimum of the space age,} compared to samples used in the past. For all empirical models, the error in ToA prediction is largest for slower CMEs. Given these uncertainties and apparent insensitivity of empirical models to deprojected speeds, we will not consider them further.

Instead, we focus on models that make certain assumptions about the CME shape and/or direction to estimate the 3D speed from heliographic image data. Models using single-viewpoint heliographic imaging data in this way are Fixed-$\phi$ \citep{1999JGR...10424739S, 2011JASTP..73.1201R}, harmonic mean  \citep{2010SoPh..267..411L} and self-similar expansion \citep{2012ApJ...750...23D}. These models use geometric arguments to derive the longitude and speed of the CME assuming that these values are constant throughout the range of the observations. The Fixed-$\phi$ method simplifies the CME to a single point (or rather a very narrow LOS extension). The harmonic mean method simplifies the CME to a circle which intersects the Sun and CME front.  The self-similar expansion method is an extension of the harmonic-mean method in that the CME is no longer anchored at the Sun but is expanding with a constant angular extent. \citet{2012SoPh..279..497L} compared the predicted ToA from the Fixed-$\phi$ and harmonic mean methods for 20 CMEs which impacted \emph{STEREO}. They found ToA errors of $\pm$33 hours for the Fixed-$\phi$ and $\pm$20 hours for the harmonic mean method both with 95\%  {error margin}. The self-similar expansion method has not been applied to CME data, at this point.

 {\citet{2006JGRA..111.4105H} used Solar Mass Ejection Imager (SMEI) heliospheric imaging data to predict the ToA of 15 CMEs in 2003 and 2004. Their results are within a range of -24 to 20 hours for all CMEs.  \citet{2010SpWea...807004H} used a 3D model to predict the ToA of three CMEs also using SMEI  data at many different elongations of the front. Their best predictions were within an hour of the CME ToA.}

Methods such as triangulation \citep{2010ApJ...710L..82L,  2011JASTP..73.1173L, 2013ApJ...769...45L} and constrained harmonic mean \citep{2010ApJ...715..493L} have been developed to take advantage of the stereoscopic data from \emph{STEREO}-SECCHI. \citet{2010ApJ...710L..82L} was able to predict the arrival of the front within 12 hours for their single CME. \citet{2013ApJ...769...45L} used both triangulation and the constrained harmonic mean method to study the kinematics of three Earth-impacting CMEs. The constrained harmonic mean method gave the best results with an error between   {-2.3 to  8.4} hours in the ToA of the CME driven shock. 

Another approach is to use the \emph{in situ} detection of the CME to constrain the imaging observations \citep{2009ApJ...694..707W, 2012SoPh..276..293R, 2011ApJ...743..101T}. \citet{2009ApJ...694..707W} used a  {multiple-function} fit to the HT data to describe the kinematics of a CME at different heliocentric distances. \citet{2012SoPh..276..293R} and \citet{2011ApJ...743..101T} have used spline fits to derive the velocity and acceleration profiles of the CMEs studied in these papers. This approach may provide some insight into the CME kinematics but cannot be used for operational space weather predictions because the ToA is no longer a free variable.

In this paper, we attempt a more operational approach based on a sample of nine Earth-impacting CMEs. We use the GCS model to fit the multi-viewpoint observations from SECCHI and LASCO  {similar to \citet{2012SoPh..279..477K}. However, unlike \citet{2012SoPh..279..477K}, we extend our fits  into the heliospheric observations as far as $\sim$1~AU, in some cases.} We then fit the derived 3D positions using a variety of models, such as constant speed or accelerating profiles, restricting the fits to certain HT ranges, taking into account the geometry of the impact and finally comparing with the \emph{in situ} measurements. Our aim is to find the simplest and most reliable model for a set of HT observations than can lead to better operational ToA predictions.

\section{Observations of Earth-impacting CMEs}
Our primary data comes from the coronagraphic and heliospheric imaging  observations of \emph{STEREO}-SECCHI and \emph{SOHO}-LASCO from March 2010 to June 2011. This data set allows us to continuously track Earth-impacting CMEs from 2 or 3 viewpoints at all times.  { \emph{STEREO} is comprised of two spacecraft with nearly identical instrumentation; the \emph{STEREO}-Ahead (STA) spacecraft orbits slightly faster than the Earth and the \emph{STEREO}-Behind (STB) spacecraft slightly slower. The two spacecraft separate from Earth at a rate of 22.5$^o$ per year since their launch on 25 October 2006.} In this study, we use \emph{STEREO}-SECCHI observations from the outer coronagraph, COR2, which has a FOV from 2.5 to 15 $R_\odot$ \citep{2008SSRv..136...67H}. SECCHI also includes two heliospheric imagers (HI-1, HI-2) which are similar to coronagraphs but have no occulter and a FOV off-pointed from the center of the Sun. The heliospheric imagers  {view the Sun-Earth line from opposite sides of the heliosphere}. HI-1 and HI-2 have square  {FOVs} centered on the elliptic plane from 15 to 84 $R_\odot$ (20$^o$)  and 66 $R_\odot$ to 1~AU (70$^o$), respectively \citep{2008SSRv..136...67H}. We also use the data from the \emph{SOHO}-LASCO  C2 (FOV 2.2--6 $R_\odot$) and C3 (FOV 3.8--32 $R_\odot$) coronagraphs \citep{1995SoPh..162..357B}.

When studying CMEs, especially Earth-impacting CMEs, it is advantageous to combine the data from LASCO and SECCHI since LASCO has a head-on view of the CME and provides a view of the extent of the CME while SECCHI has a side view and  {provides} information on the location of the CME front. During the time period of our study,  {March} 2010 - June 2011, the \emph{STEREO} spacecraft were separated from each other by 132$^o$ to 190$^o$. On 1 March 2010, STB and STA were -71$^\circ$ and 66$^o$, respectively, from Earth. The spacecraft reached opposition on 6 February 2011 and began moving closer to each other on the far side of the Sun. On 30 June 2011, STB and STA were -93$^o$ and 97$^o$ from Earth, respectively. In this configuration, an Earth-directed CME appears on the West limb in STB and on the East limb in STA.  

\end{multicols}
\begin{table}
\begin{center}
\caption{Studied CME}
\label{table1}
\begin{tabular}{ccccccccc}
\toprule
	&	\multicolumn{2}{c}{LASCO Detection} &	Halo or	&	 Lon	&	\multicolumn{2}{c}{\emph{Wind} Detection} &	Velocity	&	Detection	\\
CME 	&	 Date 	&	 Time (UT)	&	Partial	&	(deg)	&	Date	&	Time (UT)	&	 ($km s^{-1}$)	&	Type	\\ \hline
1	&	19-Mar-2010	&	 10:30	&		&	27	&	23-Mar-2010	&	23:02	&	284	&	CME	\\
2	&	03-Apr-2010	&	 10:33	&	H	&	6	&	05-Apr-2010	&	06:43	&	755	&	MC	\\
3	&	08-Apr-2010	&	 01:31	&	PH	&	-2	&	11-Apr-2010	&	11:59	&	430	&	MC	\\
4	&	16-Jun-2010	&	 14:54	&	PH	&	-18	&	20-Jun-2010	&	23:59	&	400	&	CME	\\
5	&	11-Sep-2010	&	 02:00	&	PH	&	-21	&	14-Sep-2010	&	14:24	&	368	&	MC	\\
6	&	26-Oct-2010	&	 01:36	&		&	22	&	31-Oct-2010	&	04:48	&	365	&	MC	\\
7	&	15-Feb-2011	&	 02:24	&	H	&	2	&	18-Feb-2011	&	00:00	&	510	&	MC	\\
8	&	25-Mar-2011\tablenotemark{a}	&	 14:36	&	PH\tablenotemark{b}	&	-27	&	29-Mar-2011	&	14:38	&	378	&	MC	\\
9	&	2-Jun-2011&	8:12	&	H	&	-22	&	04-Jun-2011	&	00:00	&	482	&	CME	\\ 
\bottomrule
\end{tabular}
\tablenotetext{a}{The CME was listed as two events in the \emph{SOHO} LASCO CME Catalog.}
\tablenotetext{b}{Second detection.}
\end{center}
\end{table}
\begin{multicols}{2}

To determine the ToA and speed of the CME at Earth, we use the \emph{in situ} plasma data from the \emph{Wind} spacecraft. The \emph{Wind} spacecraft, like \emph{SOHO}, orbits the L-1 Lagrange point and is ideally situated for monitoring  {near-Earth} space weather. In this study, we will use data from the \emph{Wind} Magnetic Field Investigation  (MFI; \citet{1995SSRv...71..207L}) and Solar Wind Experiment (SWE; \citet{1995SSRv...71...55O}). The MFI instrument is a triaxial magnetometer which provides the magnitude and direction of the solar wind's magnetic field. The SWE instrument provides the density, velocity and temperature of the ions of the solar wind. We use the magnetic field data to confirm the passage of a CME-like magnetic structure, an increase in magnetic field and smooth rotation in one of the field components \citep{2000GeoRL..27.3591C}. With the data from plasma instrument, we can determine the ToA and velocity of a CME to compare with our results derived from the imaging data. 

\subsection{Description of CME Event Sample}
We analyze nine Earth-impacting CMEs observed between March 2010 and June 2011 in both imaging and \emph{in situ} data.  {This time range corresponds to the rising phase of Solar Cycle 24 and is quite advantageous. CMEs during this period are more energetic but not so numerous as to result in many CME-CME interaction which confuse measurements of individual features.} The CMEs are identified in Table \ref{table1} by the date and time of their first appearance in the LASCO C2 coronagraph taken from the \emph{SOHO} LASCO CME Catalog (\url{http://cdaw.gsfc.nasa.gov/CME_list}, \citealt{2004JGRA..10907105Y}).  {We denote each CME with a number in chronological order in Table \ref{table1}. We will refer to the CMEs by these numbers, throughout this paper.} In Table \ref{table1}, we also list whether the CME was identified as a halo (H) or partial halo (PH) in the catalog. Despite all nine CMEs being Earth-directed, only three were identified as halos and four were identified as partial halos. Therefore, a CME can impact the Earth even if it is not identified as a partial halo in the LASCO catalog. With complete imaging coverage of the Sun-Earth line, we are able to show that all the studied CMEs are detected at Earth in the \emph{Wind} spacecraft data. The Heliocentric Earth Ecliptic (HEE) longitude of the CME derived from GSC model fitting is listed in  {column 5} of Table \ref{table1}.

We search the SECCHI data set beginning in January 2009,  when the spacecraft were separated by $\sim$88.5$^o$ and ending in June 2011.  To be included in our sample, the CMEs  must be observed in all the imaging data (SECCHI, LASCO, eight instruments in total) without a significant period ($< $ 1 hour) of missing data. The CME must be easily tracked between instruments. Thus the structure  of the CME had to be visible out to nearly the edge of the FOV of each instrument (with the exception of SECCHI HI-2). Due to the effects of Thomson scattering, the CME emission is dimmest from the LASCO viewpoint for Earth-impacting CMEs. Thus the visibility of the CME in the LASCO coronagraphs is usually the limiting factor for selection. To ensure we properly fit the CME envelope, we rejected any CMEs that expanded outside the upper or lower edges of the HI-1 FOV.  These restrictions are severe and eliminate many CMEs from study but are required for robust fitting of the GCS model. 

To identify the CME region in the \emph{Wind} data, we used the criteria of \citet{2005AnGeo..23.2687L} automatic detection technique. The technique was developed to detect potential magnetic clouds (MC) in the data based on the definition from \citet{1981JGR....86.6673B}. The technique can also  {identify possible} CMEs in the \emph{in situ} data. The detection requirements for a  {potential MC} are higher than for a CME detection. The minimum requirements for  {potential} CME detection are; the proton plasma beta must be  $<\beta_p>  \leq 0.3$, the field directions must change smoothly, and these two conditions must persist continuously for a minimum of eight hours. For  {possible MC} detection, a period of data must meet the minimum criteria above and have (i) a high average magnetic field strength (B $>$ 7~nT), (ii) a low proton thermal velocity (v$_{th}$=30 $km s^{-1}$) and, (iii) a minimum change in the magnetic field latitude ($\Delta \theta = 35 ^o$). All nine CMEs meet the minimum detection criteria of \citet{2005AnGeo..23.2687L}; seven of them also met the criteria for  {MC} detection.  In Table \ref{table1}, we list the \emph{in situ} detection type for each CME. The detection type only indicates if the \emph{in situ} data met the outlined criteria. To determine the presence of a MC  {or a MC-like structure} in the data further analysis would be needed \citep{2007SoPh..242..159W}.

In Table \ref{table1}, we list the time when the CME is detected at the \emph{Wind} spacecraft. There is no consensus in the literature as to which parameter of the \emph{in situ} data marks the arrival of the CME (see the discussion in \citealt{2003GeoRL..30.2232G} and \citealt{2003GeoRL..30.2233C}).  Since CMEs are large structures which can persist in the \emph{Wind} data for days, the selection of the CME arrival criteria can affect the ToA by several hours. Two parameters commonly used for the \emph{in situ} ToA of a CME are the   {time of the peak magnetic field intensity} of the shock or the beginning of the  {MC}. To properly compare the imaging to \emph{in situ} data, we determine the ToA of the CME at \emph{Wind} based on the density since it is the common physical parameter between \emph{in situ} and imaging measurements.  We do not use the peak of the shock magnetic field, if present, since it arrives before the CME density front.  Similarly, we do not use the arrival of  {MC} because it occurs after the CME density front.  Therefore, we propose that the ToA of the density increase is the most appropriate for comparison to the imaging data. In Table \ref{table2}, we list the duration of the density front, the time between the density increase and the region of low plasma beta, and the mean of the velocity detected by \emph{Wind} during the passage of the density increase.

\section{Graduated Cylindrical Shell (GCS) Model}

To locate the CME front in 3D space from the imaging data, we use the GCS model. The graduated cylindrical shell model (GCS) was developed by \citet{2006ApJ...652..763T, 2009SoPh..256..111T}. It is a forward modeling method for estimating the 3D properties and position of CMEs in white-light observations.  Unlike the methods discussed earlier \citep{1999JGR...10424739S, 2011JASTP..73.1201R, 2010ApJ...710L..82L, 2011JASTP..73.1173L, 2010SoPh..267..411L, 2012SoPh..tmp...77M} that use only the front of the CME, the GCS model is a complete 3D reconstruction of the CME envelope. Other such 3D reconstruction models as well as the GCS model are reviewed in \citet{2010AnGeo..28..203M}.
 
The GCS modeling software allows the user to fit a geometric representation of the CME envelope to all simultaneous imaging observations. The shape of the GCS model is designed to mimic that of a cylindrical magnetic flux rope. The CME is described by a curved hallow body with a circular cross-section connected by two conical legs anchored at the Sun's centers. It is important to note that the GCS model is purely geometric and does not provide any information about the magnetic field. Complete details of the model geometry can be found in  \citet{2011ApJS..194...33T}. 

\begin{figure}
\noindent\includegraphics[width=\textwidth]{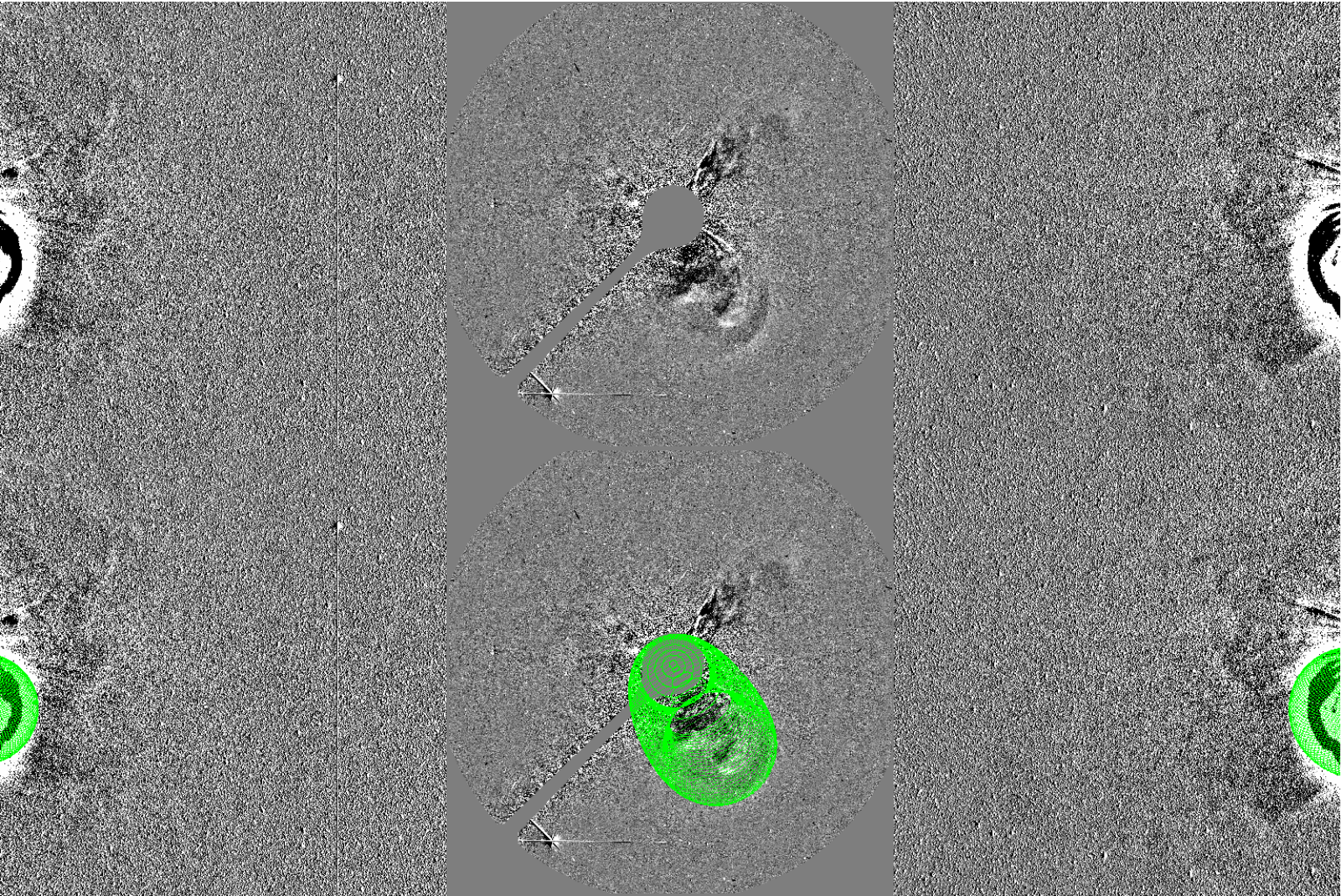}
\caption{A sample of the remote sensing data used in the study. The panels are data from STA HI-1, LASCO C3 and STB HI-1 from left to right. The data in the top and bottom panels are the same. The images in the bottom panel have been over plotted with the GCS model. The GCS model is represented by a grid of points on the surface of the model.}
\label{GCS_example}
\end{figure}

The model is fit by overplotting the projection of the cylindrical shell structure onto each image. The observer then adjusts six parameters of the model until a best visual fit with the data is achieved. The model is positioned using the longitude, latitude  and the rotation parameters. The origin of the model remains fixed at the center of the Sun. The size of the flux rope model is controlled using three parameters which define the apex height, footpoint separation and the radius of the shell.   Figure \ref{GCS_example} shows simultaneous images from  three viewpoints, STA and STB HI-1 and LASCO C3, as well as the GCS model fit to the data.  In each image, the model is projected onto the plane of the image using a grid of points (green) that represent the surface of the model. 

\subsection{Application of the GCS Model to Remote Sensing Data}

We fit the GCS model to all available images from all nine CMEs starting at the CME's first appearance in the SECCHI COR2 and LASCO C2 FOVs until the SECCHI HI-2 FOV. When the CME is visible in the LASCO data, we use all three viewpoints to make the GCS fit. The LASCO viewpoint is essential for a robust fit because the projection of an Earth-directed CME is usually quite symmetric between STA and STB.  The LASCO viewpoint can give essential information about the orientation and dimensions of the CME that is ambiguous in the SECCHI data for Earth-directed CMEs \citep{2011ApJ...733L..23V}.

Once the front of the  {CME is no longer} visible in the LASCO FOV, we must make some assumptions about its evolution. We assume that it expands self-similarly. This assumption is implemented by keeping constant all parameters of the GCS model except height. We believe self-similar expansion is a good assumption, since for most CMEs the model parameters vary only slightly when fitted using the LASCO view. A notable exception, CME 4 has a rapid change in its rotation angle in the LASCO FOV \citep{2011ApJ...733L..23V}. The effects of the rotation on the GCS model fit to this CME are discussed in \citet{2012JGRA..117.6106N}.

\end{multicols}
\begin{multicols}{2}

The GCS fitting provides measurements of various physical aspects of the CME, such as size, direction, orientation, etc. In this paper, we concentrate our analysis only on the measurements of the CME 3D height versus time (HT).  {In Figures \ref{kinematic_insitu0}-\ref{kinematic_insitu2}}, the HT measurements and \emph{in situ} data are plotted on the same time axis for each of the nine CMEs. The HT data are plotted in the  {top panel for each CME} with plus signs. We fit the GCS model at a maximum height of 211 R$_\odot$ (0.98~AU) for CME 2. The average maximum height for all the studied CMEs is 179 R$_\odot$ (0.83~AU). 
The bottom {three panels for each CME in Figures \ref{kinematic_insitu0}-\ref{kinematic_insitu2}} show the magnetic field magnitude, proton density and proton  {velocity measured \emph{in situ}} from the \emph{Wind} spacecraft.  {The first vertical green dashed line} marks the ToA of the density increase.  The second green dashed line is the backend of the density front and the beginning of the low beta plasma and smooth magnetic field rotation. The mean of the plasma velocity is also plotted  {as a horizontal green solid line in each bottom panel}. We will discuss the fits to the HT data in section~4.

\subsection{Error Estimation in Stereoscopic Localization}

To properly assess the various HT fitting methods for deriving the CME velocity and  {extrapolating} the ToA, we need to assign an error to our height measurements.  \citet{2009SoPh..256..111T} estimated the error associated with the six GCS model parameters when applied to a CME in the SECCHI COR2 views only. They found an error of $\pm 0.48$ R$_\odot$ in the height. Since we are using LASCO data in addition to the SECCHI COR2 data, we consider the errors from \citet{2009SoPh..256..111T} as an upper limit 
\begin{thinfigure}
\noindent\includegraphics[width=\linewidth]{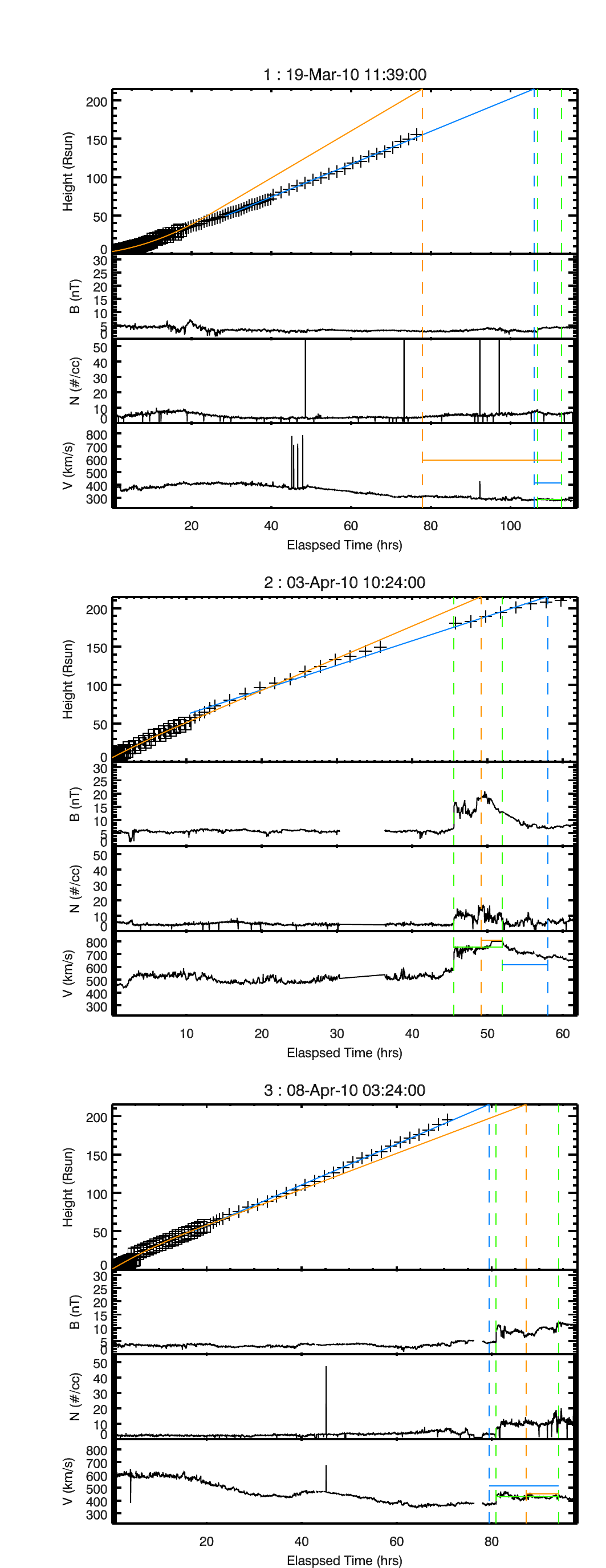}
\captionof{figure}{ HT measurements and \emph{in situ} data plotted on the same temporal axis. The HT data are plotted in the top panel with plus signs. The bottom three panels show the magnetic field magnitude, proton density and proton velocity \emph{in situ} data from the \emph{Wind} spacecraft. The vertical green dashed line marks the width of the density increase (ToA). Fit 1 and 5 and the their velocities are plotted with blue and orange solid lines, respectively. }
\label{kinematic_insitu0}
\end{thinfigure}
\begin{thinfigure}
\noindent\includegraphics[width=\linewidth]{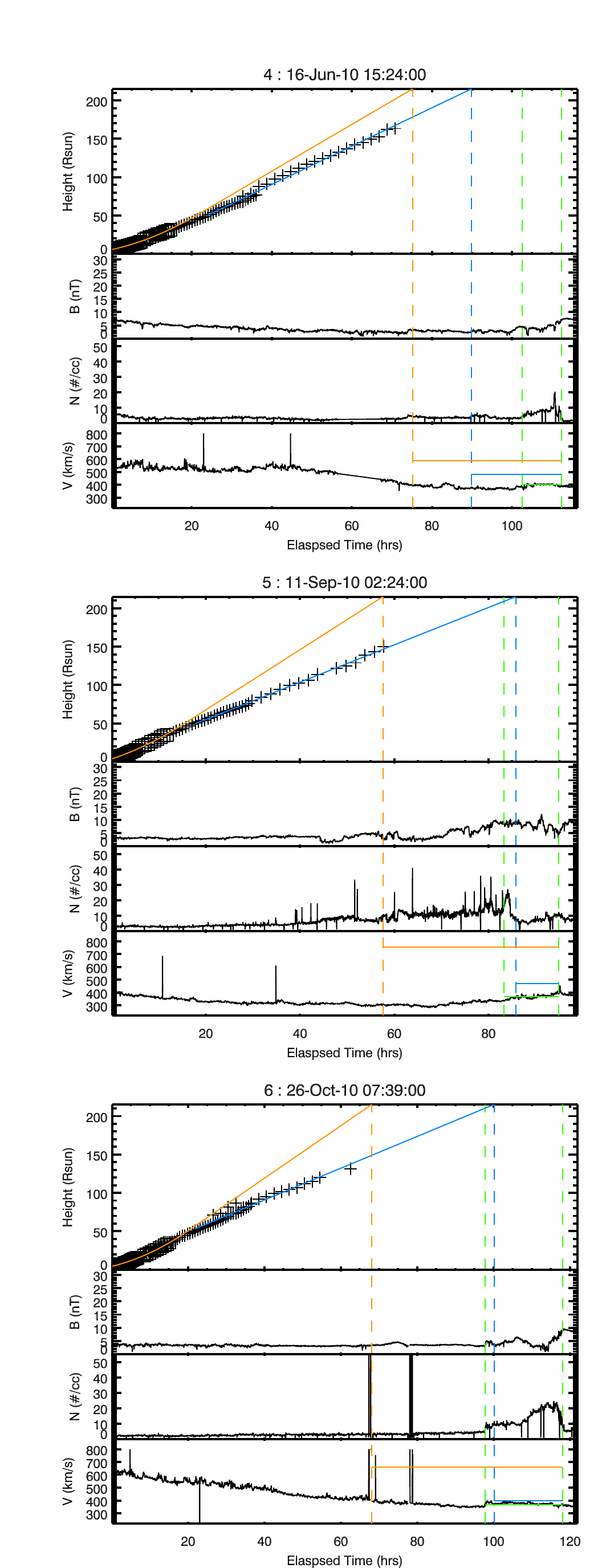}
\captionof{figure}{ Same as Figure~\ref{kinematic_insitu0} for CMEs 4-6. }
\label{kinematic_insitu1}
\end{thinfigure}
\begin{thinfigure}
\noindent\includegraphics[width=\linewidth]{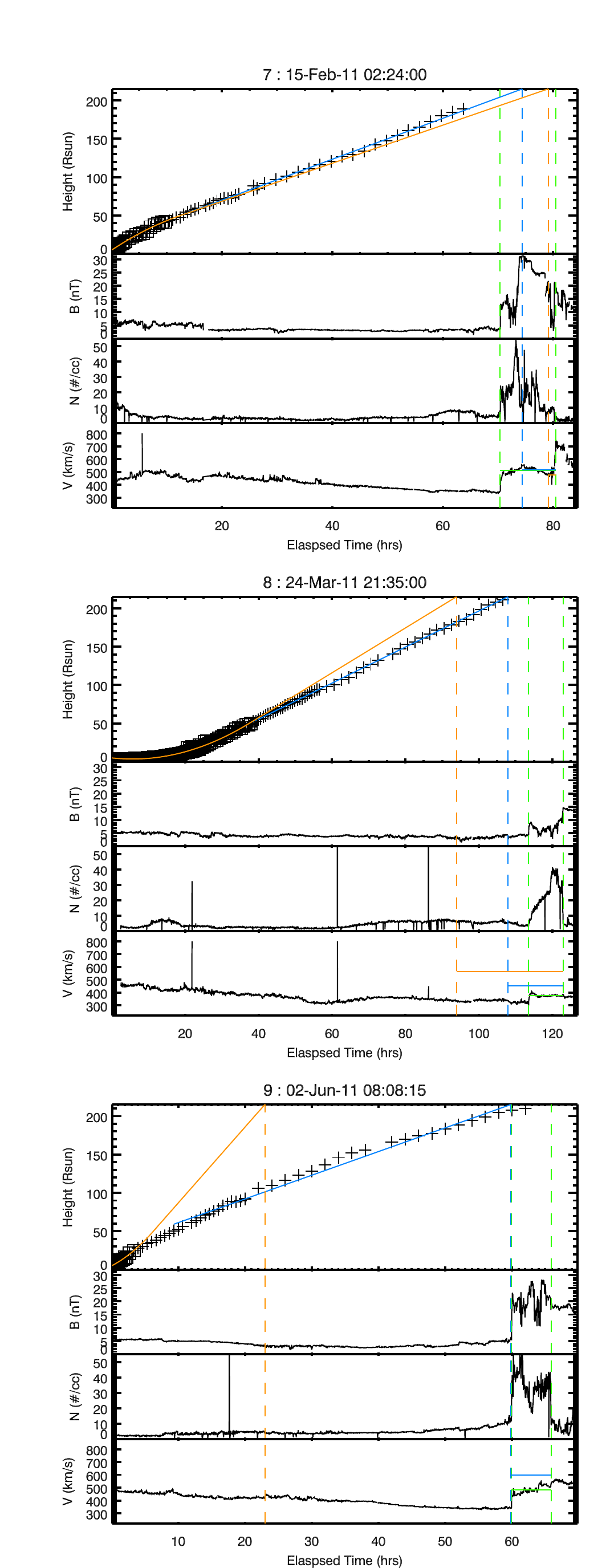}
\captionof{figure}{Same as Figure~\ref{kinematic_insitu0} for CMEs 7-9. }
\label{kinematic_insitu2}
\end{thinfigure}

\noindent for the height measurements in these FOV. Thus we need to estimate the error for heights measured in the HI images without the LASCO images. As mentioned in the previous section, once the CME in no longer visible in the LASCO FOV, we fit the GCS model to the data by only adjusting the height parameter. Thus in the HI-1 and HI-2 images, the accuracy of the GCS model fit is primarily driven by the proper localization of the CME front from the two viewpoints. 

\begin{figure}
\noindent\includegraphics[width=\textwidth]{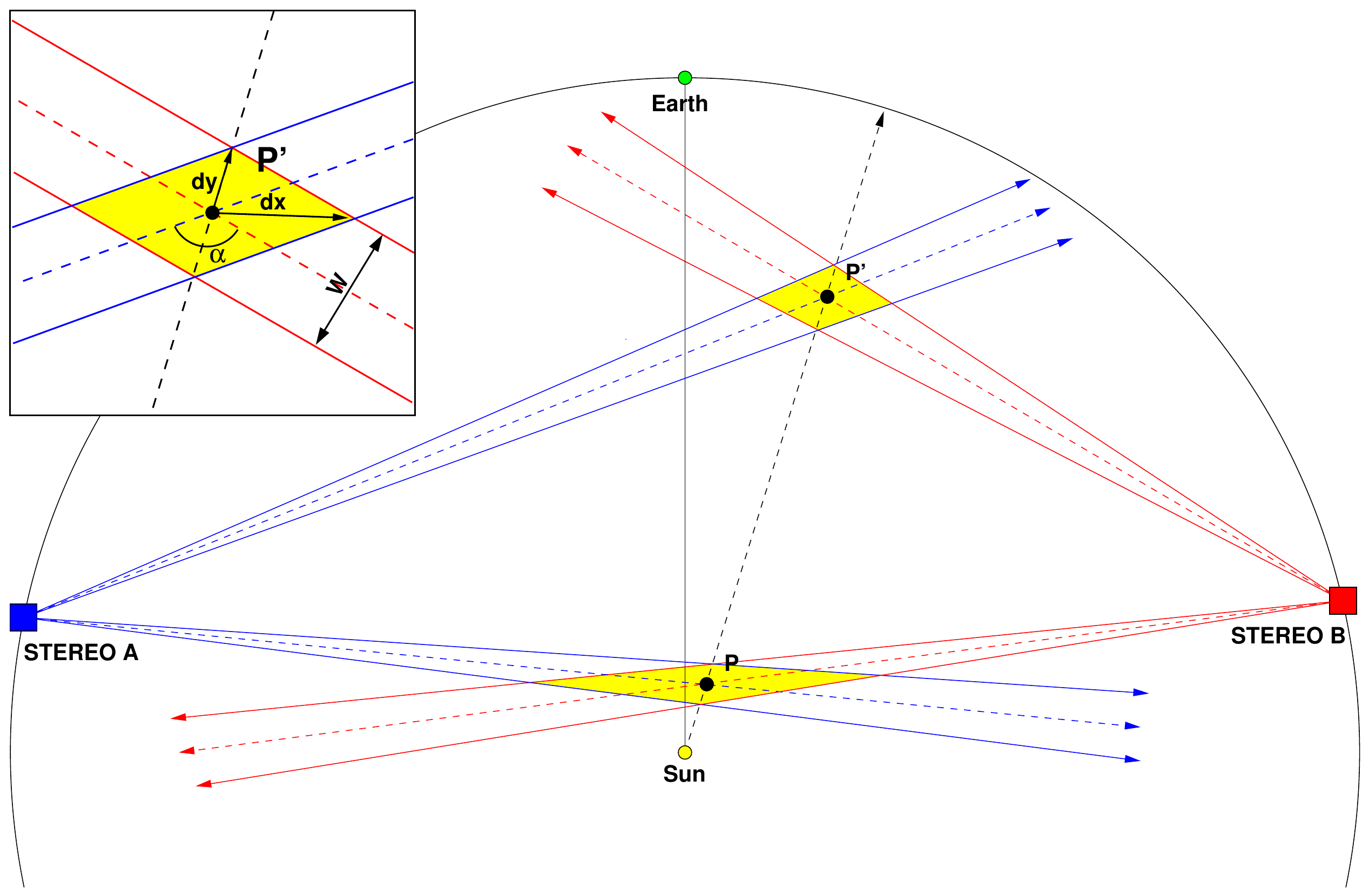}
\caption{The error in fitting of the GCS model in the HI-1 and HI-2 images can be simplified to the error in triangulating a feature in stereoscopic images. The LOS from each spacecraft is drawn with dashed lines and the cone of uncertainty is drawn with solid lines. The intersection of the uncertainty in the LOS from STA and STB creates a region of uncertainty around the feature. The insert shows a close up of the geometry for the region of uncertainty for point $P'$ assuming a long LOS.  } 
\label{error}
\end{figure}

To estimate this error, we consider the simplified problem of stereo triangulation \citep{hartley_multiple_2004}. In a digital image, there is always an error associated with the localization of a feature in the image. The error can be represented by a cone of uncertainty around the line-of-sight (LOS) from each viewpoint. In Figure \ref{error}, we represent the triangulation geometry between two points, $P$ and $P'$, near the Sun-Earth line with the \emph{STEREO} spacecraft. The LOS from each spacecraft is drawn with dashed lines and the cone of uncertainty is drawn with solid lines. The intersection of the uncertainty in the LOS from STA and STB creates a region of uncertainty around the feature. This region is a trapezoid defined by the angle between the two LOS, $\alpha$, and the uncertainty in locating the feature in the image. Thus $\alpha$ is given by
\begin{equation} 
\alpha = 2 \pi - (\theta_A + \theta_B + \varepsilon_A + \varepsilon_B ) 
\label{alpha}
\end{equation}
where $\theta_A$ and  $\theta_B$ are the longitudes of the spacecraft relative to the Sun-Earth line and  $\varepsilon_A$ and  $\varepsilon_B$ are the solar elongation of the feature in each instrument. The insert in Figure \ref{error} shows a close up of the geometry for the region of uncertainty for $P'$. Since the LOS are large and the error in locating the feature is small for the \emph{STEREO} case, we can assume that the sides of the trapezoid are  separated by a constant distance $w$. The length of the trapezoid axes, $dx$ and $dy$, are given by the equations,   
\begin{equation}
dx = \frac{w}{2 \cos{\frac{\alpha}{2}}},   \quad  dy = \frac{w}{2 \sin{\frac{\alpha}{2}}} 
\label{dxdy}
\end{equation} 
where $dx$ and $dy$ are parallel and perpendicular to the longitude of the feature, respectively. Based on our experience, we estimate the error in locating the CME front in HI-1 to be $\pm$ 5 pixels and in HI-2 is $\pm$ 10 pixels, thus, $w$ is 0.2 
\end{multicols}
\begin{multicols}{2}
\noindent $R_{\odot}$ and 1.4 $R_{\odot}$ for HI-1 and HI1-2, respectively.

Equations \ref{dxdy} require careful consideration despite their simplicity. For example,  the error $dx$ goes to infinity for $\alpha = \pi$. We can see in Figure \ref{error} that as the CME front moves between point $P$ and $P'$, that the longitude of the CME will be unconstrained. From equation \ref{alpha}, before the spacecraft reach opposition ($\theta_{A} + \theta_{B} > 2\pi$) the range of values of angle $\alpha$ includes $\alpha = \pi$. This uncertainty in the CME longitude is part of the reason why once the CME is no longer visible in the LASCO FOV, we keep the longitude of the model fixed. Since we can fit the GCS model for all the HI-1 and HI-2 images without changing the longitude, the error in the longitude must be within the minimum value of $dx$ for all measurements. If the error in the longitude is bounded by the minimum of $dx$, then the error in the height is simply $dy$ for each measurement. The maximum error in the height measurements for each CME varies between 7.4 and 12.9 $R_{\odot}$ in the HI-2 FOV. In Figures \ref{kinematic_insitu0}-\ref{kinematic_insitu2}, the error in the height is too small to be visible in the plot. The error bars for the HT measurements are shown in Figures \ref{kinematic_insitu_cme9} for the case of CME 9.

\section{Height and Time Data Fitting Methods}\label{fits}

To find the best HT fitting procedure for predicting the ToA and velocity of the CME at Earth, we explore six methods that assume various kinematic profiles for the CME front. It is not possible to measure the front height all the way to Earth for all the CMEs in our sample. The six fitting methods are described below in approximately the order of their complexity. We assign a color to each fit type which is used throughout.

\paragraph{Fit 1 - Linear (blue)} We fit a first-order polynomial to the HT measurements above a height of 50$R_\odot$ (0.23~AU). We selected the lower cutoff of 50$R_\odot$ because for most of the CMEs the HT measurements appear to be approximately linear after this height. Also 50$R_\odot$ is approximately the mean height at which the CME front is no longer visible in the LASCO data. Although the LASCO C3 FOV is 32$R_\odot$, the 3D front height for Earth-directed CME usually reaches 50$R_\odot$ within the image. Also we remind the reader that we assume self-similar expansion of the CMEs after the CME front is no longer visible in LASCO. Thus the longitude of the GSC model is fixed after 50$R_\odot$. We extrapolate the linear fit to 1~AU to find the ToA and velocity at Earth.

\paragraph{Fit 2 - Quadratic (purple)} We fit a second-order polynomial to the HT measurements above a height of 50$R_\odot$. While most of the CMEs appear to be well described by Fit 1, some of the CMEs, notably CMEs 2 and 9, have an obviously curved HT profile. This fit assumes that the CME continues to Earth with a constant acceleration. We extrapolate the function and take the first derivative at 1~AU to find the ToA and velocity at Earth.

\paragraph{Fit 3 - Multiple Polynomials (red)} We fit all available HT measurements for a given event with multiple first- and second-order polynomial functions for different time ranges. The HT measurements are fit by an initial first-order polynomial and then two second-order polynomials. The boundaries of the three functions are determined by the best fit while keeping the function and its first derivative continuous. We extrapolate the ToA by assuming a constant velocity after the final data point, again keeping the velocity continuous. This multi-function polynomial fit method is similar to that used by \citet{2009ApJ...694..707W, 2009SoPh..259..163W} to fit the kinematics of two CMEs observed in \emph{STEREO}. However, \citet{2009ApJ...694..707W, 2009SoPh..259..163W} used the ToA of the CME as a final data point for their fit.

\paragraph{Fit 4 - Spline (magenta)} We fit all HT measurements with a ridged spline. Again, we extrapolate the ToA by assuming the CME continues with a constant velocity after the final data point. The shape of the ridged spline fit is similar to Fit 3. These two methods provide similar velocity profiles. The spline fit velocity is, however, a smoothly varying curve throughout the CME trajectory which seems more physical than the velocity profiles from Fit 3 which are piecewise continuous with a discontinuous acceleration. This fit is similar to the method used by \citet{2012SoPh..276..293R} and \citet{2011ApJ...743..101T}.

\paragraph{Fit 5 - LASCO FOV (orange)} With this fit we try to compare coronagraphic analyses of the past against the heliospheric data available with \emph{STEREO}. We fit only those data points where the CME was visible in the LASCO images which is the opposite approach of Fits 1 and 2  {where we use HT measurements} after the CME front leaves the LASCO FOV.  We fit the LASCO measurements with a second-order polynomial.  We then extrapolate the ToA using a first-order polynomial with the velocity derived from the final LASCO data point. This method is similar to \citet{2012SoPh..279..477K}. However, we use a simple linear extrapolation instead of the  {empirical propagration} models of \citet{2000GeoRL..27..145G, 2001JGR...10629207G}.

\paragraph{Fit 6 - Geometric Correction (light blue)}  With this fit, we attempt to take into account the effect of the curvature of the CME front on the ToA and velocity. So we use the height of the GCS model along the Sun-Earth line instead of the apex height. These heights take into account the curvature of the GCS model front and the longitude of the CME propagation. As an example, Figure \ref{geo} shows an ecliptic cut through all GCS model fits for CME 8 where the central line of the plot is the Sun-Earth line and the dashed line is the longitudinal direction of the model. In Table \ref{table1}, we
\begin{thinfigure}
\noindent\includegraphics[width=\linewidth]{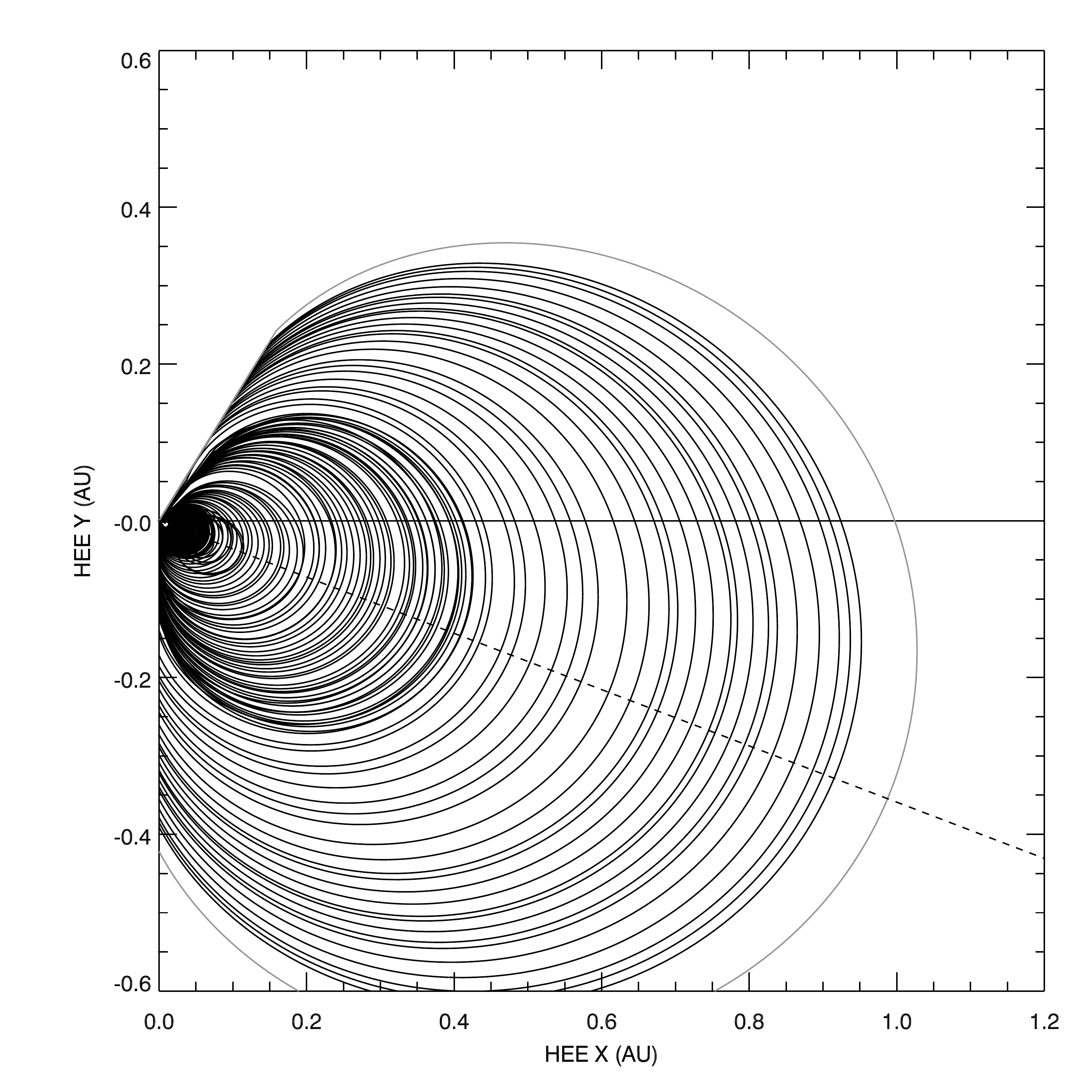}
\captionof{figure}{Ecliptic cut through the GCS model fits for CME 8 where the central line of the plot is the Sun-Earth line and the dashed line is the longitudinal direction of the model. }
\label{geo}
\end{thinfigure}
\noindent list the HEE longitude from each fit GCS. Obviously, the front height along the Sun-Earth line is less then the apex height. Thus the curvature of the CME front delays the arrival of the CME and reduces the velocity. We fit these curvature corrected HT data in the same way as Fit 1.

\begin{figure}
\begin{center}
\noindent\includegraphics[width=28pc]{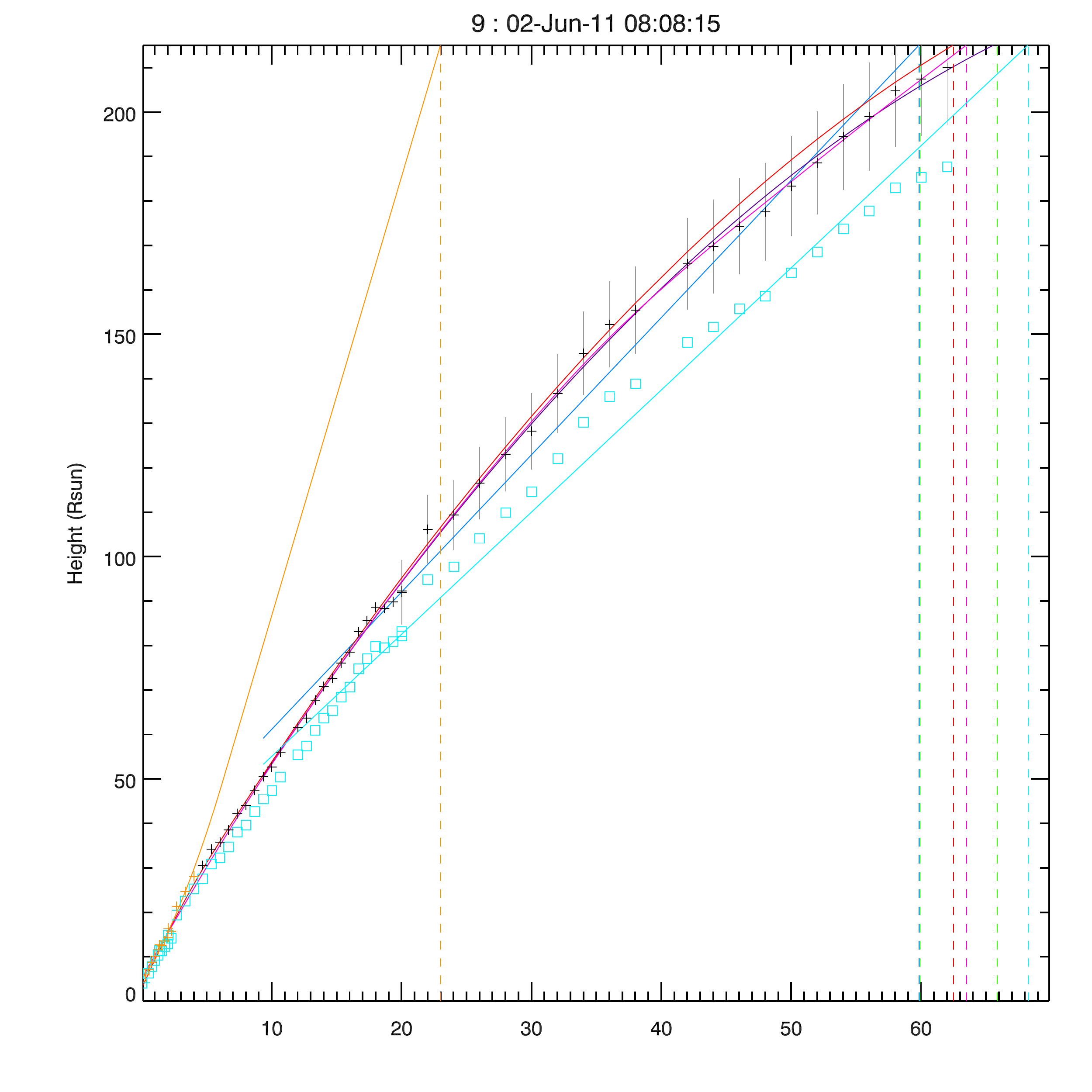}
\caption{Comparison of the six HT fitting methods for CME 9. The green dashed lines mark the time of \emph{in situ} density front. Fits 1 (blue), 2 (purple), 3 (red), 4 (magenta), 5 (orange), and 6 (light blue) and their ToA are plotted with solid and vertical dashed lines, respectively. Black crosses represent the deprojected HT measurements and light blue squares represent the same points corrected for the front curvature as derived from the GCS fit. See Section~4 for details.}
\label{kinematic_insitu_cme9}
\end{center}
\end{figure}

In Figure \ref{kinematic_insitu_cme9}, we have plotted all the fit methods for the HT measurements of CME 9. The CME 9 HT measurements of the apex are plotted with black crosses. The error for each measurement is plotted in gray. Fits 1 (blue), 2 (purple), 3 (red), 4 (magenta), 5 (orange), and 6 (light blue) and their ToA are plotted with solid and vertical dashed lines, respectively. The green dashed lines mark the time of the \emph{in situ} density front.  The light blue squares represent the HT measurements corrected for the front curvature (Fit 6) as derived from the GCS fit. In Figures \ref{kinematic_insitu0}-\ref{kinematic_insitu2}, we have plotted in the top panel Fit 1 and Fit 5 for each of the CMEs. Again the solid and vertical dashed lines represent the fits and ToA, respectively.

\section{Results}

To quantify the accuracy of the various HT fits in predicting the ToA and CME velocity at 1~AU, we calculate the difference $\Delta$T = ToA$_{predicted}$-ToA$_{\emph{Wind}}$. A negative $\Delta$T implies an early arrival and conversely, a positive $\Delta$T implies a late arrival. The $\Delta$T in hours are listed in Table \ref{table2} for each fitting method. In the first column, we list the duration of the \emph{in situ} density front in hours. In Figures \ref{kinematic_insitu0}-\ref{kinematic_insitu2}, the boundaries of the \emph{in situ} density front are marked with vertical dashed green lines. The velocity listed in Table \ref{table1}, is the mean of the measured proton velocity during the passage of the \emph{in situ} density front. The \emph{Wind} proton velocity is plotted in the bottom panels of Figures \ref{kinematic_insitu0}-\ref{kinematic_insitu2}. The mean velocity, listed in Table \ref{table1}, is plotted over the \emph{Wind} measurements with a horizontal solid green line between the dashed lines of the density front. We calculate the velocity error by finding the difference between the predicted velocity and the mean of the plasma velocity within the \emph{in situ} density front ($\Delta$V = V$_{predicted}$-$\overline{V}_{\emph{Wind}}$). In Table \ref{table2}, we list the range of the measured \emph{in situ} velocities. We have included the duration and velocity variability of the \emph{in situ} density front in our discussion because they may provide a sense of scale for the prediction errors.

\begin{figure}
\noindent\includegraphics[width=\textwidth]{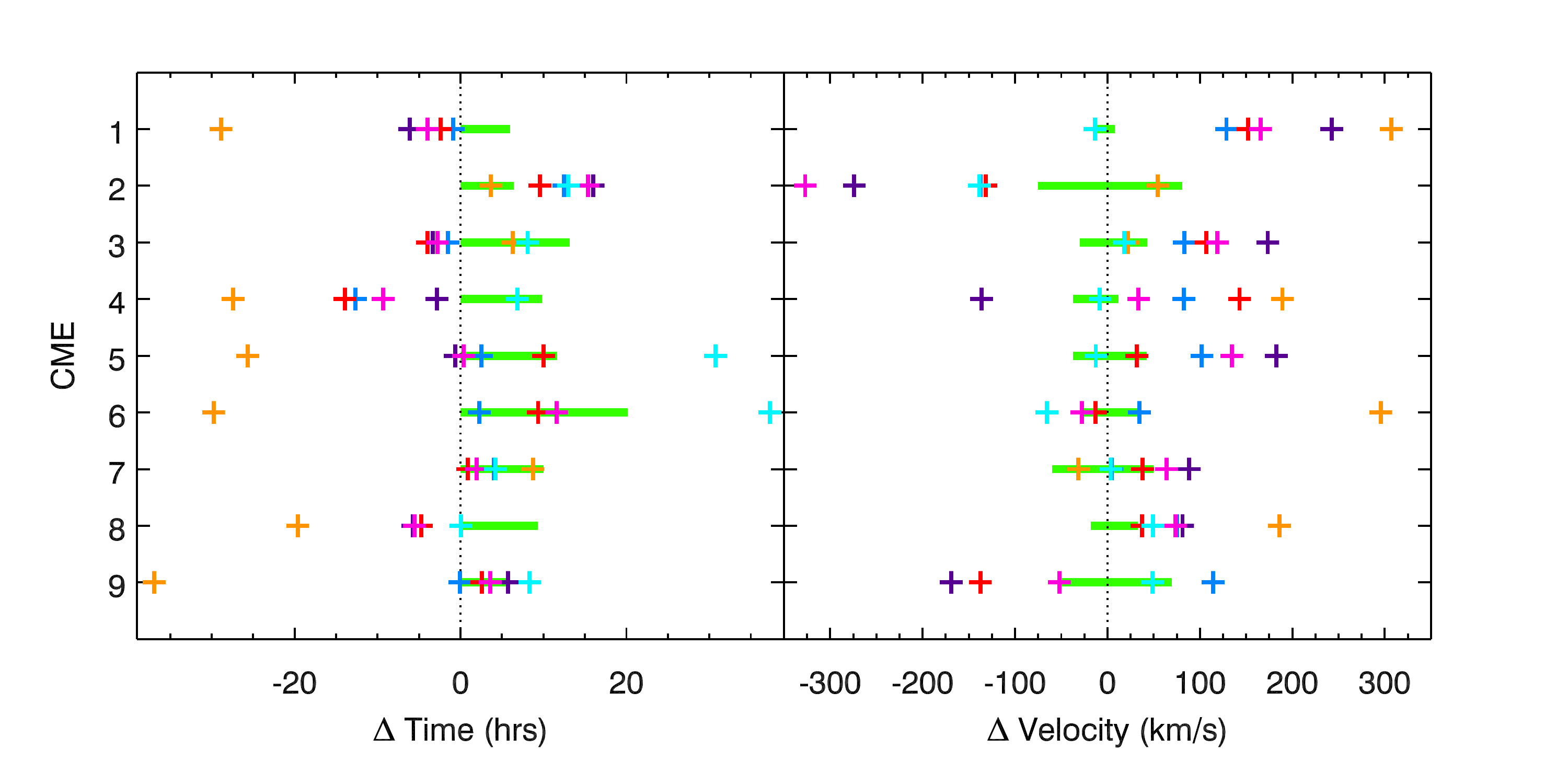}
\caption{A visual representation of the results in Table \ref{table2}. The $\Delta$T for each fit method has been plotted with plus signs in the left panel by CME number. The green line represents the duration of the CME density front. In the right panel, we have plotted $\Delta$V. The green lines represent the range of velocities detected \emph{in situ} within the density front. The results for the various fits (described in Section~4) are plotted in the following color scheme: Fit 1 (blue), Fit 2 (purple), Fit 3(red), Fit 4 (magenta), Fit 5 (orange), Fit 6 (light blue).}
\label{double}
\end{figure}

We visually represent the results from Table \ref{table2} in Figure \ref{double}. In the left panel, the $\Delta$T for each fit method is plotted with plus signs by CME number on the vertical axis. The results for the various fits follow the color code in section \ref{fits}. The green line represents the duration of the CME \emph{in situ} density front. In the right panel, we plot $\Delta$V using the same scheme. The green lines in the right panel represent the range of velocities measured within the \emph{in situ} density front.

For Fit 1, the $\Delta$T is within $\pm$ 6 hours, for seven of the CMEs. For 6 out of 9 events, the predicted ToA are either 2 hours before or within the density front. The two events (CMEs 2, 4) with  $\Delta$T$\pm$13 hour are possibly violating the self-similar expansion  {assumption} \citep{2012JGRA..117.6106N, 2011ApJ...729...70W}. It is unclear how the violation of this assumption could affect the ToA,  {furthermore}, the $\Delta$T error is  {in the opposite} sense for these two events.  The CME 2 ToA is late while CME 4 
\end{multicols}
\begin{multicols}{2}
\noindent is early. The predicted velocities from Fit 1 do not compare as well as the ToA. For only two CMEs (6 and 7), $\Delta$V is within $\pm$ 50 $km s^{-1}$ of the mean \emph{in situ} velocity.  For four of the CMEs (1, 2, 5 and 9) the $\Delta$V is greater than $\pm$ 100  $km s^{-1}$. Almost all the predicted velocities are too fast with the exception of CME 2. Clearly, all CMEs in our sample decelerate on the way to 1~AU.

The increased complexity of Fit 2 (quadratic), improves the $\Delta$T for CMEs 4, 5, and 7. Yet, the improvements to the ToA of CMEs 5 and 7 are trivial and only vary the $\Delta$T of the CME within the density front. The $\Delta$T of CME 4 is improved significantly from -12.74 to -2.94 hours. We cannot predict the ToA for CME 6 because the quadratic fit fails to intersect with 1~AU, \emph{i.e}, the CME does not make it to the Earth. The ToA for the remainder of the events is not improved with Fit 2.  This is true even for CMEs 2 and 9 which are not fit well with a constant velocity and hence Fit 2 was expected to improve $\Delta$T.  Overall, $\Delta$V is also not improved with Fit 2. Only two of the CMEs are within $\pm$ 100 $km s^{-1}$ of the mean \emph{in situ} velocity.  While the ToA of CME 4 is significantly improved with Fit 2, the predicted velocity is worse. Clearly, the quadratic fit overestimates the CME deceleration to 1~AU.

\end{multicols}
\begin{table}
\begin{center}
\caption{Error in Predicted Arrival and Velocity at Earth}
\begin{tabular}{c|cccccccc}
\toprule
	&	CME	&	Duration\tablenotemark{1}	&	Fit 1	&	Fit 2	&	Fit 3	&	Fit 4	&	Fit 5	&	Fit 6	\\ \hline
\multirow{9}{*}{\begin{sideways}$\Delta$T (hrs)\end{sideways}}																	
	&	1	&	6.00		&	-0.94	&	-6.17	&	-2.47	&	-4.07	&	-28.93	&	56.42	\\
	&	2	&	6.42		&	12.41	&	15.90	&	9.52		&	15.28	&	3.59		&	13.00	\\
	&	3	&	13.17	&	-1.58	&	-3.41	&	-4.03	&	-2.86	&	6.21		&	8.09	\\
	&	4	&	9.83		&	-12.74	&	-2.94	&	-13.97	&	-9.39	&	-27.45	&	6.83	\\
	&	5	&	11.67	&	2.47		&	-0.70	&	9.97		&	0.29		&	-25.69	&	30.76	\\
	&	6	&	20.17	&	2.18		&			&	9.30		&	11.48	&	-29.82	&	37.32	\\
	&	7	&	10.03	&	3.97		&	1.87		&	0.83		&	1.90		&	8.66		&	4.23	\\
	&	8	&	9.33		&	-5.69	&	-5.81	&	-4.81	&	-5.64	&	-19.68	&	0.04	\\
	&	9	&	5.95		&	-0.10	&	5.60		&	2.50		&	3.53		&	-36.92	&	8.34	\\ \hline
\multirow{9}{*}{\begin{sideways}$\Delta$V ($km s^{-1}$)\end{sideways}}																	
	&		&	Velocity Range\tablenotemark{2}	&		&		&		&		&		&		\\ \hline
	&	1	&	10	&	129	&	243	&	153	&	166	&	308	&	-13	\\
	&	2	&	78	&	-137	&	-273	&	-131	&	-326	&	55	&	-138	\\
	&	3	&	37	&	84	&	174	&	107	&	120	&	23	&	18	\\
	&	4	&	25	&	83	&	-136	&	143	&	34	&	190	&	-8	\\
	&	5	&	40	&	102	&	183	&	32	&	135	&	390	&	-13	\\
	&	6	&	31	&	35	&		&	-13	&	-27	&	296	&	-65	\\
	&	7	&	55	&	6	&	89	&	39	&	65	&	-31	&	4	\\
	&	8	&	25	&	76	&	82	&	38	&	74	&	187	&	50	\\
	&	9	&	61	&	115	&	-169	&	-138	&	-52	&	1426	&	49	\\ 
\bottomrule		
\end{tabular}
\label{table2}
\tablenotetext{1}{The duration of the CME density front.} 
\tablenotetext{2}{The absolute range of in situ speeds detected within the CME density front.} 
\end{center}
\end{table}
\begin{multicols}{2}

Fit 3 does not improve the ToA predictions despite having more free parameters than the pervious fits. Only the ToA for CME 7 is improved over Fits 1 and 2.  For only four CMEs, $\Delta$V is within $\pm$ 100 $km s^{-1}$ of the mean \emph{in situ} velocity. Similarly, Fit 4 with the most free parameters fails to provide an overall improvement in the predictions. 

The most important finding from this exercise may be the disappointing performance of Fit 5. Similar to \citet{2012SoPh..279..477K}, we  are investigating whether accurate 3D HT measurements in coronagraphic FOVs can be used to reliably predict the ToA of CMEs. Our results and \citet{2012SoPh..279..477K} suggest that restricting the measurements to these heights dramatically increases the ToA error compared to using the inner heliospheric measurements. Our fit uses the fewest HT measurements but these measurements are based on images from three viewpoints and are thus the most constrained.  The $\Delta$T for only three CMEs is within $\pm$12 hour. These results should be of interest to the operational Space Weather community since most CME ToA prediction methods use measurements from LASCO coronagraph along the Sun-Earth line. For this reason, we explore this fit and the influence of the final height in the ToA accuracy in the next Section.

Interestingly, this fit has the best prediction for the ToA of CME 2 (3.59 hours error) of all methods and leads us to two conclusions: 1) CME 2 underwent most of its kinematic evolution before $\sim 50 R_\odot$, and 2) the heliospheric measurements for this event are likely inaccurate. As we mentioned earlier, this is a peculiar event with an undetermined orientation which may not conform to the GCS model fitting. The six remaining CMEs are predicted to arrive $>14$ hours early and the predicted velocities are $> $100 $km s^{-1}$ higher than the \emph{in situ} velocities. The results from Fits 1-5 confirm
\end{multicols}
\begin{multicols}{2}
\noindent past findings that CMEs undergo significant deceleration above 50 R$_\odot$, on average. Our $\Delta$T results are similar to those of \citet{2012SoPh..279..477K}.

With Fit 6, we investigate the effect of the CME geometry predicted by the GCS model on the ToA. Since the front the GCS model, and presumably of the actual CME, is curved, the intersection of the CME with Earth will be delayed relative to the 1~AU arrival of the CME apex.  \citet{2012SoPh..tmp...77M} found that for a hypothetical circular CME front, the flank can be delayed by up to 2 days compared to the apex arrival at 1~AU. Our model is a bit more realistic since the front of the GCS model is not circular but slightly oblate depending on the footpoint separation. Since all ToAs are based on the CME apex height, the geometric correction of Fit 6 can only delay the ToA. Hence, only the ToA errors for CMEs 1, 3, 4, and 8 can be improved by considering the CME front geometry. With Fit 6, the ToA of CMEs 1 and 3 are "overcorrected"; the ToA is too late. The correction lowers the $\Delta$T for CMEs 4 and 8 by 6.79 and 5.65 hours, respectively. We discuss the implications in the next section. Interesting, the geometric correction improves $\Delta$V compared to Fit 1 for all CMEs, even for CMEs 1, 3, and 9, where the correction increases $\Delta$T. The $\Delta$V error is within $\pm$100 $km s^{-1}$ for eight CMEs. For CME 2, the velocity is unchanged.
  
 \section{The Effect of Final Height in Fit 5 on the ToA Accuracy } 
  We repeat Fit 5 but instead of using the last LASCO FOV measurement as the limit for the quadratic fit, we include measurements at larger heights within the HI FOV. In Figure \ref{fit_7}, we plot the resulting $\Delta$T  versus the final height of the second-order fit. The curves trace the errors for a given CME and the best prediction for each event is highlighted with a red square. The CME number is given on the right of the plot. For most of the CMEs the fit is nearly linear and as more points are added, the function become more and more linear.

It is clear, and generally expected, that the 
\begin{thinfigure}
\noindent\includegraphics[width=\linewidth]{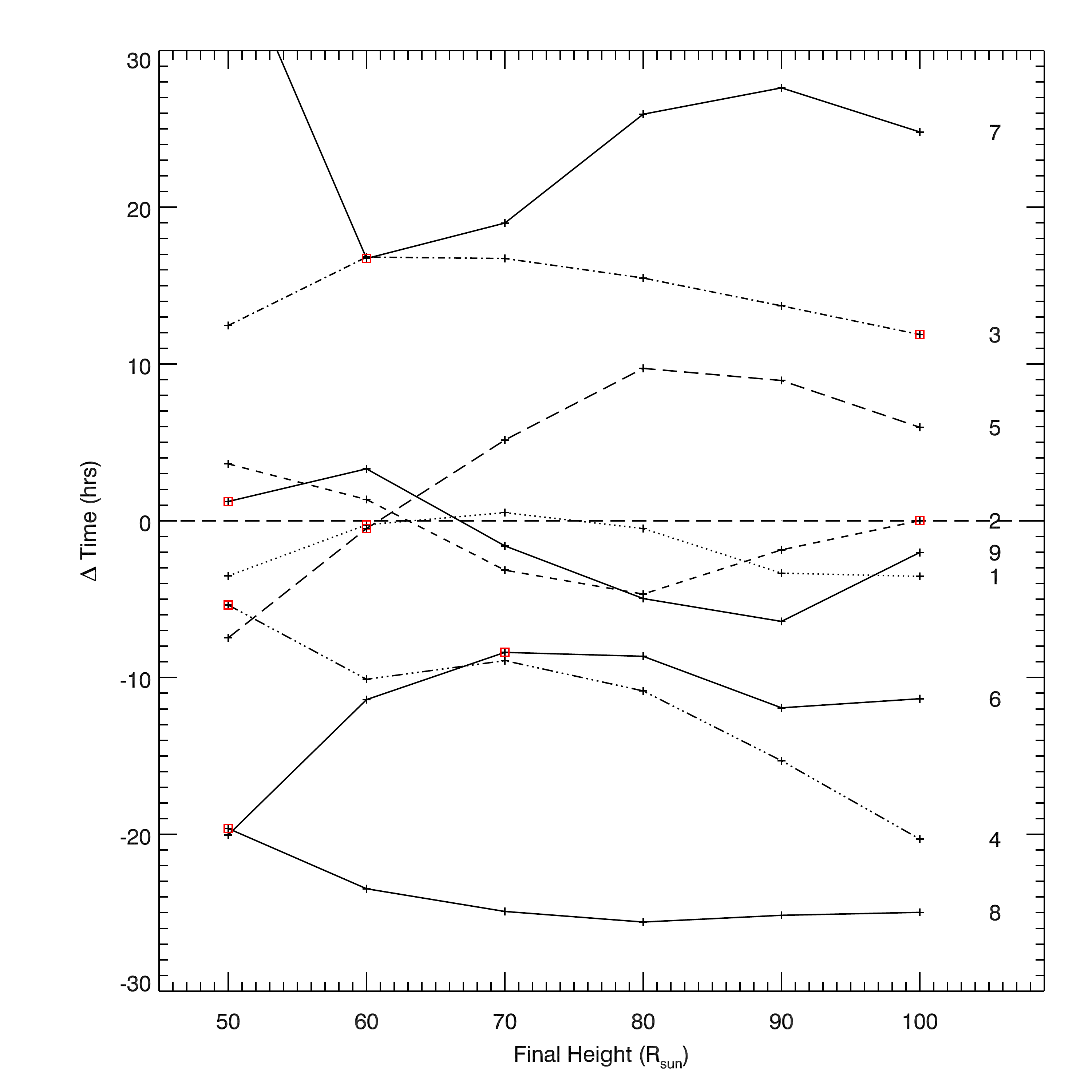}
\captionof{figure}{Time of arrival error, $\Delta$T$=ToA_{fit 5} - ToA_{\emph{Wind}}$, for Fit 5 as a function of the final height used for the fit. The curves trace the error for a given CME and the best result is shown by the red square. The event number is shown on the right end of the corresponding curve. }
\label{fit_7}
\end{thinfigure}
\noindent addition of HT measurements beyond 50 R$_\odot$ improve the ToA accuracy, sometimes considerably (ie., by 40 hours for CME 9). Interestingly, it seems that most of the gain lies in just extending the measurements to 60 R$_\odot$. Additional heights do not improve the ToA or can even make it worse (e.g., CME 4). However, this improvement does not occur for the events with the best ToAs in Fit 5.  In the case of CME 7, the additional HT measurements decrease the ToA accuracy threefold. If we ignore CMEs 4 and 8 for the moment, we see that the addition of higher HT points tends to result in later arrival times; namely, it gives slower velocities at the final point used for the quadratic fit. This is another indication that CMEs  {decelerate} above 50 R$_\odot$. However, the lower velocity bias strongly affects  the events that have already undergone the majority of their deceleration (events with $\Delta T >0$, CMEs 3, 5, 7). We do not have an obvious explanation for this at the moment. Larger sample studies are needed.

However, we can reach a couple of interesting conclusions from this exercise: (1) ToA predictions can be improved considerably with a few HT measurements in the HI FOV ($>50 R_\odot$) especially for events without strong deceleration within the LASCO FOV range. (2) ToA predictions for strongly decelerating events may be better if based on HT measurements below $50 R_\odot$. (3) There does not seem to be a ``standard'' distance range where CMEs undergo most of their deceleration, as may be suggested by the multi-polynomial plots in \citet{2009ApJ...694..707W}, for example. CMEs 3, 5, 7 seem to have decelerated by 50$R_\odot$ and to have picked up speed after this height; CMEs 1, 6, 9 seem to decelerate in the $50-60 R_\odot$ range while CME 2 or 5 seem to propagate more or less with a constant speed. 

\section{Discussion}

In this paper, we investigate methods for predicting the ToA of Earth-impacting CMEs based on de-projected HT measurements from multi-viewpoint coronagraphic and heliospheric images. From the comparison of six methods, we conclude that a simple linear fit of the HT measurements above 50 R$_\odot$ can significantly reduce the ToA error. The predicted ToA from the linear fit (Fit 1) is within $\pm$6 hours of the arrival of the  density front at the \emph{Wind} spacecraft for 78\% of CMEs. This result is a 9 hour improvement over the results of \citet{2001JGR...10629207G}  that reports an accuracy of  $\pm$15 hour for 72\% of CMEs studied.  If we include all events in our study, we can predict the arrival of CMEs at Earth with $\pm$13 hours which is almost a half day improvement over the $\pm$24 hour window with a 95\%  {error margin} previously reported in \citet{2005AnGeo..23.1033S}. Our results are also an improvement over the Fixed-$\phi$ and harmonic mean methods, $\pm$33  and $\pm$20 hours, respectively, which use heliospheric data without taking advantage of the two \emph{STEREO} viewpoints \citep{2012SoPh..279..497L}.  Even our worst case results are a significant improvement in predicting CME ToA. Therefore,  \textsl{deprojected HT measurements using images of the CME front obtained from outside the Sun-Earth line can improve the accuracy of the ToA prediction of Earth-impacting CMEs by a half day compared to single-view coronagraphic observations obtained along the Sun-Earth line.}

The CMEs with the poorest ToA results (2 and 4) are peculiar. They may violate the self-similar expansion assumption used to fit the GCS model to the HI-1 and HI-2 images.  \citet{2012JGRA..117.6106N} found that CME 4 is rotating between 0.5~AU and 1~AU and that its appearance is subject to considerable projection effects. The CME 2 orientation is ambiguous despite being the subject of several studies. \citet{2011ApJ...729...70W}, for example, found that the cross section of the CME is significantly elliptical irrespective of the actual orientation. An elliptical cross-section may indicate that the expansion of the CME was not self-similar; rotation is also likely. In any case, the heliospheric HT measurements for this CME are suspect as it is the only event with an improved ToA from Fit 5. Given the small sample of CMEs, and the even smaller number of discrepancies, we cannot reach a firm conclusion on whether CME rotation or other projection effects may be responsible for the poor ToA predictions. 

\begin{figure}
\centering
\noindent\includegraphics[width=28pc]{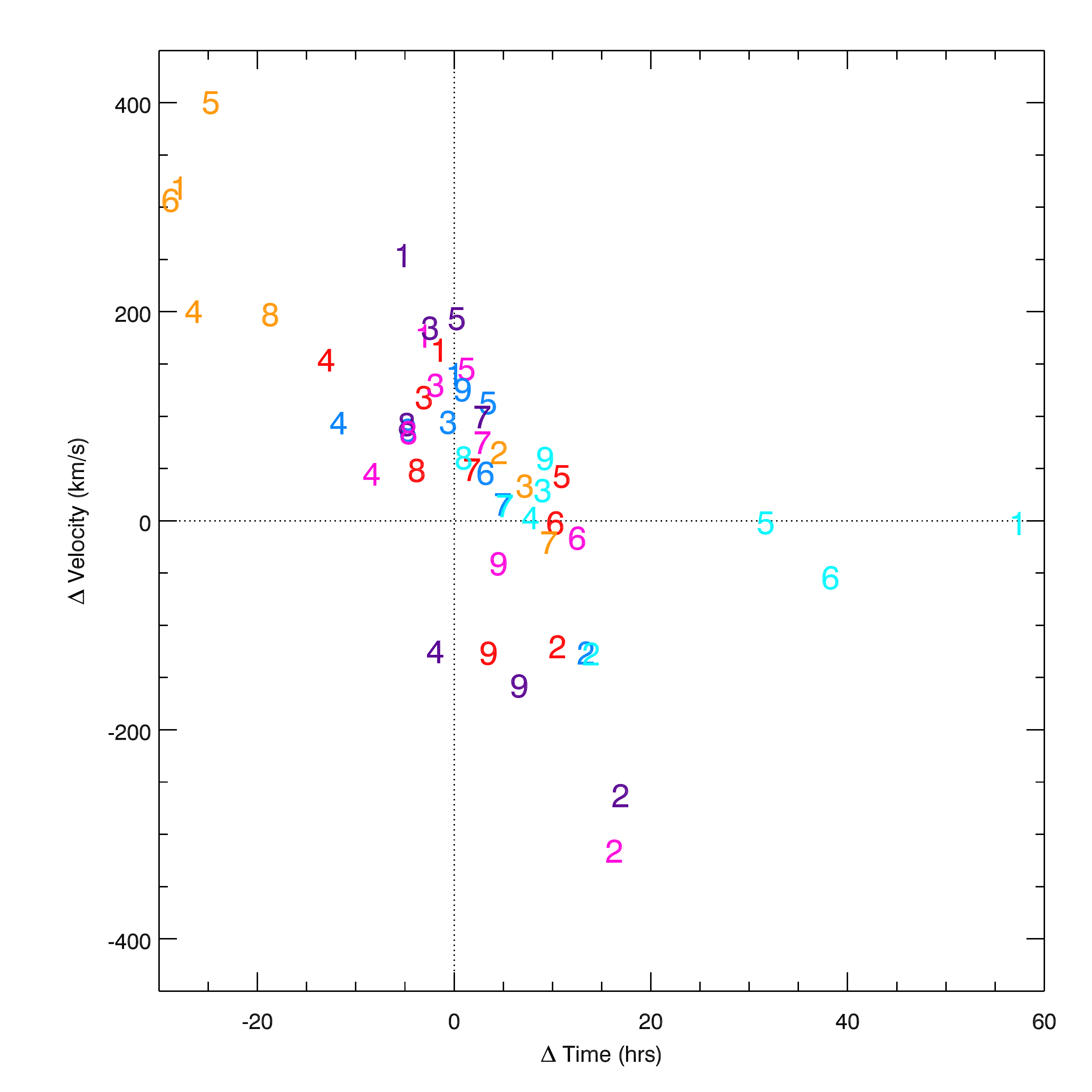}
\caption{The error in the arrival time, $\Delta$T versus the error in the 1~AU velocity, $\Delta$V. The color scheme is the same as in Figure~\ref{double}. Counterintuitively, there is no obvious correlation between the two variables, with the exception of Fit 5 (orange). There is a slight tendency to overestimate the arrival velocity by about 50 $km s^{-1}$.}
\label{scatter}
\end{figure}

We are not aware of any previous studies of the CME ToA that report the predicted velocity at 1~AU as well. We think that this is a serious omission, since a reliable prediction of the CME velocity at Earth can, in turn, provide reliable estimates of the CME ram pressure and hence help predict one more geo-effective parameter. We also use the predicted velocity as a diagnostic of our fit methods.  Since the distance traveled by the CMEs is fixed, we would assume a correlation with $\Delta$T and $\Delta$V. In other words, if the fitted velocity is too fast, we would expect the CME to arrive early and vice versa.  {In Figure \ref{scatter}, we have plotted $\Delta$T versus $\Delta$V where the results are plotted using the CME number and the color scheme from section \ref{fits}. It }is clear that while $\Delta$T is evenly distributed around zero (with the exception of Fit 5), $\Delta$V is largely positive. More precisely, $\overline{\Delta T} = 1.1$ hours and  $\overline{\Delta V}$ = 53 $km s^{-2}$.  There is no obvious trend or correlation, between $\Delta$T and $\Delta$V, within a particular fit or among the fitting methods with the exception of Fit 5. For Fit 5, the faster velocities are somewhat correlated with early ToA, as one would expect.  Fit 6 has the smallest velocity error but it has the three largest ToA errors.  The geometric correction of Fit 6 systematically decreases the predicted velocity, as expected, but it does not increase the ToA accuracy. We conclude that  \emph{a linear fit to the HT measurements above 50$R_\odot$ is sufficient for predicting the ToA but fails to capture the true kinematics of the CME. } 

This result would seem to suggest that the CMEs are traveling at a constant speed between 50 $R_\odot$ and 1~AU.  However, closer analysis of our of results does not support this claim.  First, if the CME reached a constant velocity by 50$R_\odot$, we would expect the results from Fit 5 (LASCO FOV only) to be as accurate as Fit 1. But Fit 5 results in early arrivals which implies that the velocity derived at 50$R_\odot$ with the quadratic fit is too high and hence the velocity of the CME must decrease after 50 $R_\odot$. This deceleration, however, must occur very gradually otherwise Fit 2 (quadratic) would perform better than Fit 1 (linear).  It is well known that the velocities measured in the LASCO FOV have a broader range compared the velocities at Earth which converge around the average solar wind speed \citep{2000GeoRL..27..145G}. However, it is not known at what heights CMEs reach a constant velocity. We assert a CME should reach a constant velocity only after its velocity matches the ambient solar wind velocity. If there is a difference in the velocity of the CME and the ambient solar wind, the CME will be effected by a drag force \citep{2002JGRA..107.1019V}. Six of our CMEs (2, 3, 4, 6, 7, 8, 9) exhibit an abrupt increase in the \emph{in situ} velocity coincident with the density front. Thus, they are still traveling faster than the solar wind and are still decelerating. For the two CMEs that are traveling with the solar wind velocity (1 and 5), the $\Delta$V from Fit 5 is 308 $km s^{-2}$ and 390 $km s^{-2}$, respectively. Therefore, these CME decelerated sometime after 50$R_{\odot}$ but before reaching 1~AU and did so smoothly since the linear fit gives the best ToA for those events. We conclude that \emph{ all CMEs in our sample are decelerated between 50$R_{\odot}$ and 1~AU.}

\end{multicols}
\begin{multicols}{2}

We assume that the primary cause of the CME deceleration is the drag force due to the interaction with the ambient solar wind.  While the drag force could also accelerate a CME between between 50$R_{\odot}$ and 1~AU, we did not measure such a CME in our sample. While the drag force varies as $|V_{CME}-V_{SW}|^2$, the effect on the velocity would not be quadratic. The drag force is degenerate; as the velocity of the CME decreases so does the drag force. Thus the deceleration due to the drag would be very gradual and occur smoothly as we see in our HT measurements. Also the transition of the CME into equilibrium with the solar wind would also occur smoothly. This would explain why the HT profiles are not well fit by Fit 2 (quadratic) and Fit 3 (multiple polynomials) but are better represented by Fit 1 (linear). We believe that Fit 4 (rigid spline) failed because there is too much error in the HT measurements.  

But we have to reconcile our two conclusions: (1) A linear fit to the HT data is the best method for predicting the ToA; (2) All measured CME are decelerating.  The obvious suggestion is that the linear fit provides the mean velocity of the gradually decelerating CME front between 50 $R_\odot$ and 1~AU. This also explains the systematic overestimation of the CME velocity with Fit 1. The mean velocity of a gradually decelerating function will always be higher than the final velocity. Thus we somewhat alter our original conclusion. \emph{The mean velocity of a CME between 50 $R_\odot$ and 1~AU is the best parameter for predicting the ToA.}  {The linear fit is a simple method for calculating the average velocity from the HT data.}

We compare the results from Fit 1 and 6 since the HT measurements were fitted in the same way. For Fit 6 we used the height of the GCS model along the Sun-Earth line as opposed to the apex height in Fit 1 (see Figure \ref{geo}). The corrected height is less than the apex height depending on the \textsl{width\/} and \textsl{longitude\/} of the GCS model. We find that the apex height is a better predictor of the CME arrival.  We interpret this result as evidence for flattening of the CME fronts during Earth propagation. The flattening of the CME front has been theorized in the past \citep[][and references therein]{2004ApJ...600.1035R} and seems to occur in the HI-1 and HI-2 images perpendicular to the ecliptic (but see discussion in \citet{2012JGRA..117.6106N}). In the heliographic images, we do not have reliable information about the extent or curvature of the CME in the ecliptic plane. However, if the curvature of the CME was the dominant factor in the CME ToA error, we would expect the results of Fit 1 to be systematically early ($-\Delta$T). We do not see this. Only CMEs 1, 3, 6, and 8 have early predicted ToA and, therefore, could benefit from the correction. However, all four "corrected" arrivals result in much later ToA, i.e., they are overcorrected. Thus we have to assume that the front of these CMEs is not as curved the GCS model predicts. Therefore, we have indications of flattening of the CME front in the ecliptic beyond 50 R$_\odot$, for some events. Further investigations on the role of projection effects \citep[e.g,][]{2012JGRA..117.6106N} and on the proper identification of the CME substructures \citep[e.g.,][]{2013SoPh..284..179V} is needed.

\section{Conclusions}
With Fits 1 to 4,  we add complexity with each fit by increasing the number of free parameters in an attempt to capture the  {kinematics} of the CME in the heliosphere. We assume that the increased number of free parameters would result in better fits to the HT measurements and that the ToA and velocity prediction would correspondingly improve. Surprisingly, Fit 1 while having the fewest free parameters, gives the best results. We find that the best results are obtained by ignoring complex fitting functions to the full data range, even discarding the coronal observations, and fitting a simple straight line to the HT measurements above 50 R$_\odot$ only. We show that measurements close to the Sun, as those provided by coronagraphs, are not sufficiently robust for ToA predictions even if those HT measurements are deprojected somehow. Furthermore, we find that being able to follow a CME front all the way to Earth  (e.g., CMEs 2 and 8 but see CME 9 for a counterexample) does not actually improve the ToA. Correcting for the CME curvature does not improve the ToA. Imaging observations integrate along a long LOS, which becomes longer with increasing heliocentric distance. Therefore, the location of a CME feature can be subject to considerable uncertainty, including a bias towards the location of the Thompson sphere \citep{2006ApJ...642.1216V}, if the CME is undergoing rotation or other interaction with the ambient environment. Such evolution is likely to affect the derived CME longitude and its curvature.

These results have important implications for Space Weather and CME propagation studies:
\begin{enumerate}
\item A simple linear fit to deprojected HT measurements of the CME front only above 50 $R_\odot$ is sufficient to predict the ToA within $\pm 6$ hours (for 7/9 events) and the 1~AU velocity within $\pm$ 140 $km s^{-1}$.
\item Deprojected HT measurements of CMEs made using imaging from outside the Sun-Earth line can improve the Earth ToA prediction of CMEs by a half day compared to single-view coronagraphic observations along the Sun-Earth line.
\item CMEs decelerate slowly and smoothly between 50$R_{\odot}$ and 1~AU.
\item HT measurements within coronagraphs FOVs (30 $R_\odot$) even if they are deprojected, are insufficient for accurate Earth ToA or CME velocity predictions. 
\item Despite the improvements in CME size and direction, achieved using \emph{STEREO} data, there remain several open issues in the interpretation of the images such as the precise localization of the Earth-impacting part of the CME. 
\end{enumerate}


%

\paragraph{Acknowledgments}
  R.C and A.V are supported by NASA contract S-136361-Y to the Naval
  Research Laboratory. C.W. is supported by Navy ONR 6.1 program. The SECCHI data are produced by an
  international consortium of the NRL, LMSAL and NASA GSFC (USA), RAL
  and Univ.  Bham (UK), MPS (Germany), CSL (Belgium), IOTA and IAS
  (France). LASCO was constructed by a consortium of institutions:
  NRL, MPIA (Germany), LAM (France), and Univ. of Birmingham (UK). The
  LASCO CME catalog is generated and maintained at the CDAW Data
  Center by NASA and The Catholic University of America in cooperation
  with the Naval Research Laboratory.



\end{multicols}

\end{document}